\documentclass[journal]{IEEEtran}
\usepackage{graphicx,cite,epsfig,amssymb,amsmath}
\usepackage{epstopdf}

\usepackage{pifont}
\usepackage{stmaryrd}
\usepackage{bbding}
\usepackage{algorithm}
\usepackage{algorithmic}
\usepackage{mathrsfs}
\usepackage{latexsym}
\usepackage{indentfirst}
\usepackage{fmtcount}
\usepackage{cite}
\usepackage{float}
\usepackage{morefloats}

\usepackage{caption}
\usepackage{subcaption}

\usepackage{color}
\usepackage{mathtools}
\usepackage{xfrac}

\usepackage{array,colortbl,multirow,multicol,booktabs,ctable,bigstrut}

\usepackage[
top    = 0.75in,
bottom = 0.75in,
left   = 0.75in,
right  = 0.75in]{geometry}

\newtheorem{prop}{Proposition}
\newtheorem{remark}{Remark}
\newtheorem{notation}{Notation}

\begin{document}
%
\title{A Flip-Syndrome-List Polar Decoder Architecture for Ultra-Low-Latency Communications}
%
%
%

\author{
    \IEEEauthorblockN{Huazi~Zhang, Jiajie~Tong, Rong~Li, Pengcheng~Qiu, Yourui~Huangfu, Chen~Xu, Xianbin~Wang, Jun~Wang}\\
    \IEEEauthorblockA{Huawei Technologies Co. Ltd.}\\
    Email: \{zhanghuazi,tongjiajie,rongone.li,justin.wangjun\}@huawei.com
}
\maketitle

\begin{abstract}
We consider practical hardware implementation of Polar decoders. To reduce latency due to the serial nature of successive cancellation (SC), existing optimizations \cite{Dec:SSC,Dec:SSCL,Dec:fast_SSCL,Dec:multi-bit,Dec:ML-SSC,Dec:GoodBits,Dec:SplitReduce} improve parallelism with two approaches, i.e., multi-bit decision or reduced path splitting. In this paper, we combine the two procedures into one with an error-pattern-based architecture. It simultaneously generates a set of candidate paths for multiple bits with pre-stored patterns. For rate-1 (R1) or single parity-check (SPC) nodes, we prove that a small number of deterministic patterns are required to guarantee performance preservation. For general nodes, low-weight error patterns are indexed by syndrome in a look-up table and retrieved in $O(1)$ time. The proposed flip-syndrome-list (FSL) decoder fully parallelizes all constituent code blocks without sacrificing performance, thus is suitable for ultra-low-latency applications. Meanwhile, two code construction optimizations are presented to further reduce complexity and improve performance, respectively.
\end{abstract}


\section{Introduction}\label{section:intro}
\subsection{Background and related works}
Polar codes \cite{Polar:Arikan,Polar:Stolte} have been selected for the fifth generation (5G) wireless standard. With state-of-the-art code construction techniques \cite{Polar:GA,Polar:PW,Polar:CA_List_Niu} and SC-List (SCL) decoding algorithm \cite{Dec:SSC,Dec:SSCL,Dec:fast_SSCL,Dec:multi-bit,Dec:ML-SSC,Dec:GoodBits,Dec:SplitReduce,Polar:List_Tal}, Polar codes demonstrate competitive performance over LDPC and Turbo codes in terms of block error rate (BLER). Beyond 5G, ultra-low decoding latency emerges as a
key requirement for applications such as autonomous driving and virtual reality. The latency of practical Polar decoders, e.g., an SC-list decoder with list size $L=8$, is relatively long due to the serial processing nature.

Continuous efforts \cite{Dec:SSC,Dec:SSCL,Dec:fast_SSCL,Dec:multi-bit,Dec:ML-SSC,Dec:GoodBits,Dec:SplitReduce} have been made to significantly reduce decoding latency. Among them, we are particularly interested in hardware implementations, which are dominant in real-world products, due to better power- and area-efficiency. According to our cross-validation, three approaches are shown to be cost-effective, yet incur no or negligible performance loss compared to the original SCL decoder, as summarized below:
\begin{enumerate}
\item Pruning on the SC decoding tree \cite{Dec:SSC} (parallelizing constituent code blocks with mult-bit decision)
\begin{itemize}
\item Rate-0 (R0), repetition (Rep) nodes \cite{Dec:SSCL,Dec:fast_SSCL}.
\item General (Gen) nodes comprised of consecutive bits \cite{Dec:multi-bit,Dec:ML-SSC}.
\end{itemize}
\item Reduce the number of path splitting
\begin{itemize}
\item Rate-1 (R1), single parity-check (SPC) nodes \cite{Dec:fast_SSCL}.
\item Do not split upon the most reliable (good) bits \cite{Dec:GoodBits,Dec:SplitReduce}.
\end{itemize}
\item Reduce the latency of list pruning
\begin{itemize}
\item Adopt bitonic sort \cite{Dec:bitonic} for efficient pruning.
\item Quick list pruning \cite{Dec:double_threshold}.
\end{itemize}
\end{enumerate}

\subsection{Motivation and our contributions}\label{section:Contribution}
It is well known that an SC decoder requires $2N-2$ time steps for a length-$N$ code \cite{Polar:Arikan}. The SC decoding factor graph reveals that, the main source of latency is the left hand side (LHS, or information bit side) of the graph. In contrast, the right hand side (RHS, or codeword side) of the graph consists of independent code blocks and already supports parallel decoding.

With the above observations, the key to low-latency decoding is to parallelize LHS processing. Existing hardware decoder designs are pioneered by \cite{Dec:SSC,Dec:SSCL,Dec:fast_SSCL,Dec:multi-bit,Dec:ML-SSC}, which view SC decoding as binary tree search, i.e., a length-$N$ code (a parent node) is recursively decomposed into two length-$N/2$ codes (child nodes). Upon reaching certain special nodes, their child nodes are not traversed \cite{Dec:SSC} and the corresponding path metrics are directly updated at the parent node \cite{Dec:SSCL}. Even though, there is still room for further optimizations:
\begin{itemize}
  \item The processing of an R1/SPC node is not fully parallel (e.g., a number of sequential path extension \& pruning are still required \cite{Dec:fast_SSCL}). A higher degree of parallelism can be exploited to further reduce latency.
  \item Optimizations (e.g., parallel processing) are applied to some special nodes (e.g., R0/Rep/SPC/R1), and the length of such blocks, denoted by $B$, is often short due to insufficient polarization. According to our measurement under typical code lengths, the main source of latency is now incurred by the general nodes whose constituent code rates are between $\frac 2 B$ and $\frac {B-2} B$.
\end{itemize}


Motivated by \cite{Dec:SSC,Dec:SSCL,Dec:fast_SSCL,Dec:multi-bit,Dec:ML-SSC}, and thanks to the recent advances in efficient list pruning \cite{Dec:bitonic,Dec:double_threshold}, we find it profitable to further improve parallelism for ultra-low-latency applications. Our contributions are summarized below:
\begin{enumerate}
  \item We propose to fully parallelize the processing of R1/SPC nodes via multi-bit hard estimation and flipping at intermediate stages. Only one-time path extension/pruning per node is required by applying a small number of flipping patterns on the raw hard estimation. Such simplification is proven to preserve performance.
  \item For general nodes, we apply flip-syndrome-list (FSL) decoding to constituent code blocks. Specifically, a small set of low-weight error patterns are pre-stored in a table indexed by syndrome. During decoding, syndrome is calculated per constituent code block. Its associated error patterns are retrieved from the syndrome table, and used for bit-flip-based sub-path generation. Similar to R1/SPC nodes, the FSL decoder narrows down the candidates for path extension, and enjoys the simplicity of a hard-input decoder. The proposed optimization is shown to incur negligible performance loss.
  \item The complexity of an FSL decoder is mainly incurred by constituent code blocks with medium rates. We propose to re-adjust the distribution of information bits in order to avoid certain constituent code rates, such that decoder complexity can be significantly reduced. We show that the performance loss can be negligible.
  \item With the FSL decoder's capability to decode arbitrary linear outer constituent codes, not necessarily Polar codes, we propose to adopt hybrid outer codes with optimized distance spectrum. The hybrid-Polar codes demonstrate better performance than the original Polar codes.
\end{enumerate}

Paper is organized as follows, Section \ref{section:Polar} introduces the fundamentals of Polar SCL decoding, Section \ref{section:FSL} provides the details of FSL decoder including R1/SPC nodes, general nodes, latency analysis and BLER performance, Section \ref{section:construction} proposes two improved code construction methods that benefit from the FSL decoder architecture, Section \ref{section:conclusion} concludes the paper.

\section{Polar Codes and SCL Decoding}\label{section:Polar}
A binary Polar code of mother code length $N=2^n$ can be defined by $\textbf{c}=\textbf{u}\textbf{G}$ and a set of information sub-channel indices $\cal I$. The information bits are assigned to sub-channels with indices in $\cal I$, i.e., $\textbf{u}_{\cal I}$, and the frozen bits (zero-valued by default) are assigned to the rest sub-channels. The Polar kernel matrix is $\textbf{G} = \textbf{F}^{\otimes n}$, where $\textbf{F} = \begin{bmatrix}1 & 0 \\ 1 & 1\end{bmatrix}$ is the kernel and $^\otimes$ denotes Kronecker power, and $\textbf{c}$ is the code word.  The transmitted BPSK symbols are $\textbf{x}_0^{N-1} = 1-2\cdot\textbf{c}_0^{N-1}$ and the received vector is $\textbf{y}_0^{N-1}$.

For completeness, the original SCL decoder \cite{Polar:List_Tal} is briefly revisited. The SC decoding factor graph of a length-$N$ Polar code consists of $N \times (\log_2 N+1)$ nodes. The row indices $i=\{0,1,\cdots,N-1\}$ denote the $N$ bit indices. The column indices $s=0,1,\cdots,\log_2 N$ denote decoding stages, with $s=0$ labeling the information bit side and $s=\log_2 N$ labeling the input LLR side (or codeword side). Each node in the factor graph can be indexed by a $(s,i)$ pair, and is associated with a soft LLR value $\alpha_{s,i}$, which is initialized by $\alpha_{\log_2 N,i}=y_i$, and a hard estimate $\beta_{s,i}$.

For all $s$ and $i$ satisfying $i\mod 2^{s+1} < 2^s$, a hardware-friendly right-to-left updating rule for $\alpha$ is:
\begin{align*}
& \alpha_{s,i} = \text{sgn}(\alpha_{s+1,i})\text{sgn}(\alpha_{s+1,i+2^s})\min(|\alpha_{s+1,i}|,|\alpha_{s+1,i+2^s}|),\\
& \alpha_{s,i+2^s} = (1-2\beta_{s,i})\alpha_{s+1,i} + \alpha_{s+1,i+2^s}.
\end{align*}

The hard estimate of the $i$-th bit is $\beta_{0,i} = \frac {1-\text{sgn}(\alpha_{0,i})} 2$.
The corresponding left-to-right updating rule for $\beta$ is:
\begin{align*}
& \beta_{s,i} = \beta_{s-1,i} \oplus \beta_{s-1,i+2^{s-1}},\\
& \beta_{s,i+2^{s-1}} = \beta_{s-1,i+2^{s-1}}.
\end{align*}

An SCL decoder with list size $L$ executes path split upon each information bit, and preserves $L$ paths with smallest path metrics (PM).
Given the $l$-th path with $\hat u_i^l$ as the $i$-th hard output bit, a hardware-friendly PM updating rule \cite{Dec:LLR_based_SCL} is
\begin{align*}
& \text{PM}_{i}^l =  \left\{\begin{array}{cc}
   \text{PM}_{i-1}^l, & if \ \hat u_i^l = \beta_{0,i}^l, \\
   \text{PM}_{i-1}^l + |\alpha_{0,i}^l|, & otherwise, \\
  \end{array} \right.
\end{align*}
where $\text{PM}_{i}^l$ denotes the path metric of the $l$-th path at bit index $i$, and $\alpha_{0,i}^l$ and $\beta_{0,i}^l$ denote its corresponding soft LLR and hard estimation, respectively.

After decoding the last bit, the first path\footnote{For CRC-aided Polar, the first path that passes CRC check is selected.} is selected as decoding output.

\section{Flip-Syndrome-List (FSL) Decoding}\label{section:FSL}
SC-based decoding of length-$N$ Polar codes requires $\log_2 N+1$ stages to propagate received signal ($s=\log_2 N$) to information bits ($s=0$). The degree of parallelism is $2^s$, i.e., reduces by half after each decoding stage.

To increase parallelism, we propose to terminate the LLR propagation at intermediate stage $s=\log_2 B$, and process all length-$B$ constituent code blocks with a hard-input decoder. The design is detailed throughout this section, where differences to existing works mainly include (i) fully parallelized processing for $B$ bits and $L$ paths, and (ii) supporting arbitrary-rate blocks rather than special ones (e.g., R0/Rep/SPC/R1).

\subsection{Multi-bit hard decision at intermediate stage}
The indices of a constituent code block is denoted by ${\cal B} \triangleq \{i,i+1,\cdots,i+B-1\}$. Once the soft LLRs at the $s$-th stage are obtained, where $s=\log_2 B$, a raw hard estimation is immediately obtained by
\begin{equation}
{\pmb \beta}_{s,{\cal B}} = \frac {1-\text{sgn}({\pmb \alpha}_{s,{\cal B}})} 2.
\end{equation}

In contrast to SCL that uses the soft LLR ${\pmb \alpha}_{s,{\cal B}}$, a constituent block decoder takes ${\pmb \beta}_{s,{\cal B}}$ as its hard input, and directly generates hard code word $\hat{\pmb \beta}_{s,{\cal B}}$ as decoded output.

The hard-input decoders for R1, SPC and general nodes will be described next in Section~\ref{section:bit_flipping} and \ref{section:syndrome_decoding}. For now, we assume such a decoder outputs a hard code word $\hat{\pmb \beta}_{s,{\cal B}}$ for each candidate path, and recover the corresponding information vector by
\begin{equation}
\hat{\textbf{u}}_{\cal B} = \hat{\pmb \beta}_{s,{\cal B}}\textbf{F}^{\otimes s}.
\end{equation}

Given the soft LLRs $\hat{\pmb \alpha}_{s,{\cal B}}$ and the recovered codeword $\hat{\pmb \beta}_{s,{\cal B}}$, the multi-bit version of PM updating rule \cite{Dec:SSCL} is
\begin{equation}\label{equ:multi_bit_PM_update}
\text{PM}_{i}^l = \text{PM}_{i-B}^l + \sum\limits_{j\in {\cal B}}\left(\left|\hat{\beta}_{s,j}^l- {\beta}_{s,j}^l\right| \left|\alpha_{s,j}^l\right|\right).
\end{equation}

The remaining updating of $\alpha$ and $\beta$ is based on the hard decision $\hat{\pmb \beta}_{s,{\cal B}}$ rather than the raw estimation ${\pmb \beta}_{s,{\cal B}}$.

\subsection{Parallelized path extension via bit flipping}\label{section:bit_flipping}
\subsubsection{Rate-1 nodes}
For an R1 node, the state-of-the-art decoding method \cite{Dec:fast_SSCL} requires $\min(L-1,B)$ times path extensions. First, the input soft LLRs ${\pmb \alpha}_{s,{\cal B}}^l$ for each list path is sorted. Then, path extensions are performed only on the $\min(L-1,B)$ least LLR positions to reduce complexity. Such simplification incurs no performance loss since additional path extensions are proven to be redundant \cite{Dec:fast_SSCL}. The searching space becomes $L\times2^{\min(L-1,B)}$, much smaller than $L\times2^{B}$ for conventional SCL \cite{Polar:List_Tal} and SSCL \cite{Dec:SSCL}. Another work \cite{Dec:fast_SCL} also proposes to reduce searching space for R1 nodes. But its candidate paths generation is LLR-dependent, thus is suitable for software implementation as suggested in \cite{Dec:fast_SCL}.

\begin{figure}
\centering
    \includegraphics[width= 0.5\textwidth]{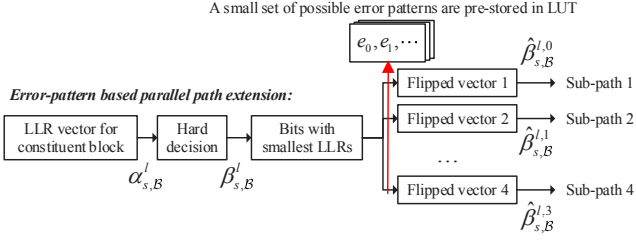}
    \caption{Error-pattern based parallel path extension.}
    \label{fig:parallel_extension}
\end{figure}
In this paper, we focus on hardware implementation and propose a parallel path extension based on pre-stored error-patterns. As shown in Fig.~\ref{fig:parallel_extension}, only one-time path extension and pruning is required for a constituent block. The optimization exploits the deterministic partial ordering of incremental path metrics within a block. Accordingly, the search for survived paths can be narrowed down to a limited set, and pre-stored in the form of error patterns in a look-up table (LUT). The LUT is shown to be very small for a practical list size $L=8$. As such, the advantages are:
\begin{itemize}
  \item $B$ bits are decoded in parallel.
  \item Sub-paths are generated in parallel.
  \item The above two procedures are combined into one.
\end{itemize}

\begin{notation}[soft/hard vectors]\label{def:sorted_alpha}
The soft LLR input of a constituent block is indexed by ascending reliability order, i.e., ${\pmb \alpha}_{s,{\cal B}}^l$ such that $|\alpha_{s,0}^l| < |\alpha_{s,1}^l| < \cdots < |\alpha_{s,B-1}^l|$ for each list path. The corresponding raw hard estimation is denoted by ${\pmb \beta}_{s,{\cal B}}^l\triangleq \left[\beta_{s,0}^l,\beta_{s,1}^l,\cdots,\beta_{s,B-1}^l\right]$.
\end{notation}

\begin{notation}[sub-paths extension]
For a constituent block with indices $\cal B$, a sub-path that extends from the $i$-bit to the $(i+B-1)$-th bit can be well defined by the blockwise decoding output. For example, the $t$-th sub-path of the $l$-th path is denoted by the vector $\hat{\pmb \beta}_{s,{\cal B}}^{l,t}$.
\end{notation}

\begin{notation}[bit-flipping]
Each vector $\hat{\pmb \beta}_{s,{\cal B}}^{l,t}$ is generated by flipping ${\pmb \beta}_{s,{\cal B}}^l$ based on an error pattern ${\pmb e}$. A single-bit-error pattern is denoted by ${\pmb e}_{p}$ if it has one at the $p$-th bit position ($p=0,1,\cdots$) and zeros otherwise.
\end{notation}

For $L=8$, we narrow down the searching space per list path from $2^{\min(L-1,B)}$ to 13 by the following proposition.
\begin{prop}\label{prop:rate_1_node}
For each path in an SCL with $L=8$, its $L$ maximum-likelihood sub-paths (i.e., with minimum incremental path metrics) fall into a deterministic set of size 13. These sub-paths can be obtained by bit flipping the original hard estimation of each list path based on the following error patterns:
\begin{equation}\label{equ:rate_1_patterns}
\hat{\pmb \beta}_{s,{\cal B}}^{l,t} = \left\{\begin{array}{cc}
    {\pmb \beta}_{s,{\cal B}}^l, & t=0; \\
    {\pmb \beta}_{s,{\cal B}}^l \oplus {\pmb e}_{t-1}, & 1\leq t\leq 7, \\
    {\pmb \beta}_{s,{\cal B}}^l \oplus {\pmb e}_{0} \oplus {\pmb e}_{1}, & t=8, \\
    {\pmb \beta}_{s,{\cal B}}^l \oplus {\pmb e}_{0} \oplus {\pmb e}_{2}, & t=9, \\
    {\pmb \beta}_{s,{\cal B}}^l \oplus {\pmb e}_{1} \oplus {\pmb e}_{2}, & t=10, \\
    {\pmb \beta}_{s,{\cal B}}^l \oplus {\pmb e}_{0} \oplus {\pmb e}_{3}, & t=11, \\
    {\pmb \beta}_{s,{\cal B}}^l \oplus {\pmb e}_{0} \oplus {\pmb e}_{1} \oplus {\pmb e}_{2}, & t=12. \\
  \end{array} \right.
\end{equation}
\end{prop}
\begin{IEEEproof}
To survive from the sub-paths of all $L$ paths, a sub-path must first survive from the sub-paths of its own parent path. That means for each parent path, we only need to consider its $L$ maximum-likelihood sub-paths. Altogether, there are at most $L^2$ sub-path to be considered.

According to \eqref{equ:multi_bit_PM_update}, the path metric penalty is received only on the flipped positions. For each sub-path and its associated error patterns, the incremental path metric is computed by
\begin{align}\label{equ:incremental_PM_of_12_patterns}
\Delta \text{PM}_{i+B-1}^{l,t} &\triangleq \text{PM}_{i+B-1}^{l,t}-\text{PM}_{i}^{l}\\ \nonumber
& = \left\{\begin{array}{cc}
    0, & t=0; \\
    |\alpha_{s,t-1}^l|, & 1\leq t\leq 7, \\
    |\alpha_{s,0}^l| + |\alpha_{s,1}^l|, & t=8, \\
    |\alpha_{s,0}^l| + |\alpha_{s,2}^l|, & t=9, \\
    |\alpha_{s,1}^l| + |\alpha_{s,2}^l|, & t=10, \\
    |\alpha_{s,0}^l| + |\alpha_{s,3}^l|, & t=11, \\
    |\alpha_{s,0}^l| + |\alpha_{s,1}^l| + |\alpha_{s,2}^l|, & t=12, \\
  \end{array} \right.
\end{align}

Since the indices of soft LLRs $|{\pmb \alpha}_{s,{\cal B}}^l|$ are ordered according to \emph{Notation}~\ref{def:sorted_alpha}, the incremental path metrics also satisfy a set of partial order, as shown in the following directed graph. The arrow ``$\rightarrow$'' denotes a ``smaller than'' relationship that can be easily verified.

{\footnotesize{
\begin{align*}
& \quad 0\\
& \ \ \downarrow \\
& |\alpha_{s,0}^l| \rightarrow |\alpha_{s,1}^l| \rightarrow |\alpha_{s,2}^l| \rightarrow |\alpha_{s,3}^l| \rightarrow |\alpha_{s,4}^l| \rightarrow |\alpha_{s,5}^l| \rightarrow |\alpha_{s,6}^l| \rightarrow \cdots \\
&  \qquad \quad \quad \swarrow \qquad \qquad \ \searrow \qquad \quad \quad \searrow\\
&  \quad |\alpha_{s,0}^l|+|\alpha_{s,1}^l| \rightarrow |\alpha_{s,0}^l|+|\alpha_{s,2}^l| \rightarrow |\alpha_{s,0}^l|+|\alpha_{s,3}^l| \rightarrow\cdots\\
&  \qquad \qquad \qquad \qquad \qquad \quad \ \downarrow \qquad \qquad \qquad \qquad \downarrow\\
&  \qquad \qquad \qquad \qquad \quad |\alpha_{s,1}^l|+|\alpha_{s,2}^l| \rightarrow |\alpha_{s,1}^l|+|\alpha_{s,3}^l| \rightarrow\cdots\\
&  \qquad \qquad \qquad \qquad \qquad \quad \ \downarrow \\
&  \qquad \qquad \qquad \qquad \quad |\alpha_{s,0}^l|+|\alpha_{s,1}^l|+|\alpha_{s,2}^l| \rightarrow \cdots
\end{align*}
}}

We prove \emph{Proposition}~\ref{prop:rate_1_node} with the above directed graph. Any node with a minimum distance to the root node ``0" larger than $L=8$ cannot survive path pruning.

First, if the 8-th smallest incremental path metric is caused by a single bit error, then it cannot be $|\alpha_{s,7}^l|$ or larger, otherwise there will be more than 8 sub-paths with incremental path metrics smaller than the 8-th one, which contradicts the assumption. The argument is obvious since there are already 8 nodes upstream of $|\alpha_{s,7}^l|$ in the directed graph.

Similarly, the 8-th smallest incremental path metric caused by two bit errors cannot be equal to or larger than $|\alpha_{s,1}^l|+|\alpha_{s,3}^l|$, because there are already more than 8 sub-paths with smaller path metrics in its upstream.

Finally, the sub-paths with incremental path metric $|\alpha_{s,0}^l|+|\alpha_{s,1}^l|+|\alpha_{s,2}^l|$ also has 8 nodes in its upstream (including itself), and any error patterns with larger incremental path metric (including the 4-bit patterns) will lead to contradiction if they are included in the surviving set.

Thus, we can reduce the tested error patterns per path to 13 with only one-time path extension without any performance loss.
\end{IEEEproof}
\begin{remark}
The bit-flipping-based path extension is mainly constituted of binary/LUT operations. The 13 error patterns are pre-stored. The resulting path metrics for all error patterns can be computed in parallel according to \eqref{equ:multi_bit_PM_update} or \eqref{equ:incremental_PM_of_12_patterns}.

The path extension and pruning are as summarized by ``$(13\rightarrow8\rightarrow 64 \rightarrow 8) \times 1$", explained as follows. For each path, the 13 error patterns lead to 13 sub-paths, among which the 8 with smallest path metrics are pre-selected $(13\rightarrow8)$. Altogether, there will be $8\times L=64$ extended paths $(8\rightarrow64)$ for the case of $L=8$. The 64 extended paths are pruned back to 8 $(64 \rightarrow 8)$. The above procedures are executed only one time. In contrast, the fast-SSCL decoder \cite{Dec:fast_SSCL} requires $L-1=7$ times path extension and pruning, i.e., $(8\rightarrow 16\rightarrow 8) \times 7$. According to Section~\ref{section:latency_analysis}, the minimum number of ``cycles'' reduces from 49 to 14 in the case of a length-16 R1 block. To avoid any misunderstanding, the ``cycles'' here captures implementation details in our fabricated ASIC \cite{ASIC}, thus should be distinguished from the ``time steps'' concept in \cite{Dec:fast_SSCL}.
\end{remark}

\begin{remark}
The proposition addresses list size $L=8$, but its idea naturally extends to all list sizes as long as the corresponding error patterns are identified. Among them, decoders with list size $L=8$ are particularly important since they are widely accepted by the industry during the 5G standardization process \cite{3GPP:Chairmannote_RAN86b}. The conclusion is drawn after extensive evaluations on the tradeoff among BLER, latency, throughput and power consumption, in which decoders with $L=8$ achieve the best overall efficiency. The tradeoff in real hardware is further verified in our implemented decoder ASIC in \cite{ASIC}.
\end{remark}

\subsubsection{SPC nodes}
For an SPC node, the state-of-the-art decoding method \cite{Dec:fast_SSCL} requires $\min(L,B)$ times path extensions. In this work, we propose only one-time path extension and reduce the searching space from $2^{\min(L,B)}$ to 13 as follows.
\begin{prop}\label{prop:SPC_node}
For SCL with $L=8$, following \emph{Notation}~\ref{def:sorted_alpha}, if the checksum of ${\pmb \beta}_{s,{\cal B}}^l$ is even, i.e., $\sum\limits_{j\in{\cal B}} {\beta}_{s,j}^l = 0$, then the $L$ surviving paths can be obtained from bit flipping each list path based on the following 13 error patterns:
\begin{equation}\label{equ:SPC_even_patterns}
\hat{\pmb \beta}_{s,{\cal B}}^{l,t} = \left\{\begin{array}{cc}
    {\pmb \beta}_{s,{\cal B}}^l, & t=0; \\
    {\pmb \beta}_{s,{\cal B}}^l \oplus {\pmb e}_{0} \oplus {\pmb e}_{t}, & 1\leq t\leq 7, \\
    {\pmb \beta}_{s,{\cal B}}^l \oplus {\pmb e}_{1} \oplus {\pmb e}_{2}, & t=8, \\
    {\pmb \beta}_{s,{\cal B}}^l \oplus {\pmb e}_{1} \oplus {\pmb e}_{3}, & t=9, \\
    {\pmb \beta}_{s,{\cal B}}^l \oplus {\pmb e}_{1} \oplus {\pmb e}_{4}, & t=10, \\
    {\pmb \beta}_{s,{\cal B}}^l \oplus {\pmb e}_{2} \oplus {\pmb e}_{3}, & t=11. \\
    {\pmb \beta}_{s,{\cal B}}^l \oplus {\pmb e}_{0} \oplus {\pmb e}_{1} \oplus {\pmb e}_{2} \oplus {\pmb e}_{3}, & t=12. \\
  \end{array} \right.
\end{equation}
Otherwise, if the checksum of ${\pmb \beta}_{s,{\cal B}}^l$ is odd, i.e., $\sum\limits_{j\in{\cal B}} {\beta}_{s,j}^l = 1$, then the $L$ surviving paths can be obtained from bit flipping each list path based on the following 13 error patterns:
\begin{equation}\label{equ:SPC_odd_patterns}
\hat{\pmb \beta}_{s,{\cal B}}^{l,t} = \left\{\begin{array}{cc}
    {\pmb \beta}_{s,{\cal B}}^l \oplus {\pmb e}_{t}, & 0\leq t\leq 7, \\
    {\pmb \beta}_{s,{\cal B}}^l \oplus {\pmb e}_{0} \oplus {\pmb e}_{1} \oplus {\pmb e}_{2}, & t=8, \\
    {\pmb \beta}_{s,{\cal B}}^l \oplus {\pmb e}_{0} \oplus {\pmb e}_{1} \oplus {\pmb e}_{3}, & t=9, \\
    {\pmb \beta}_{s,{\cal B}}^l \oplus {\pmb e}_{0} \oplus {\pmb e}_{2} \oplus {\pmb e}_{3}, & t=10, \\
    {\pmb \beta}_{s,{\cal B}}^l \oplus {\pmb e}_{1} \oplus {\pmb e}_{2} \oplus {\pmb e}_{3}, & t=11, \\
    {\pmb \beta}_{s,{\cal B}}^l \oplus {\pmb e}_{0} \oplus {\pmb e}_{1} \oplus {\pmb e}_{4}, & t=12, \\
  \end{array} \right.
\end{equation}
\end{prop}
\begin{IEEEproof}
The proof follows that of \emph{Proposition}~\ref{prop:rate_1_node}. For simplicity, we change the directed graph to Table~\ref{tab:SPC_partial_order}, where the right and lower cells are always larger than the left and upper ones.
\begin{table}
  \centering
  \caption{Partial order of path metrics for a SPC node}
{\footnotesize{
\begin{tabular}{|c|c|c|c|}
  \hline
  \multicolumn{4}{|c|}{Even checksum case for ${\pmb \beta}_{s,{\cal B}}^l$} \\ \hline
  0 & -- & -- & -- \\   \hline
  $|\alpha_{s,0}^l|+|\alpha_{s,1}^l|$ & -- & -- & -- \\  \hline
  $|\alpha_{s,0}^l|+|\alpha_{s,2}^l|$ & $|\alpha_{s,1}^l|+|\alpha_{s,2}^l|$ & -- & -- \\  \hline
  $|\alpha_{s,0}^l|+|\alpha_{s,3}^l|$ & $|\alpha_{s,1}^l|+|\alpha_{s,3}^l|$ & $|\alpha_{s,2}^l|+|\alpha_{s,3}^l|$ & $\sum\limits_{j=0}^3|\alpha_{s,i}^l|$ \\  \hline
  $|\alpha_{s,0}^l|+|\alpha_{s,4}^l|$ & $|\alpha_{s,1}^l|+|\alpha_{s,4}^l|$ & $\cdots$ & $\cdots$ \\  \hline
  $\cdots$ & $\cdots$ & $\cdots$ & $\cdots$ \\  \hline
  $|\alpha_{s,0}^l|+|\alpha_{s,7}^l|$ & $\cdots$ & $\cdots$ & $\cdots$ \\ \hline \hline
  \multicolumn{4}{|c|}{Odd checksum case for ${\pmb \beta}_{s,{\cal B}}^l$} \\ \hline
  $|\alpha_{s,0}^l|$ & -- & -- & -- \\  \hline
  $|\alpha_{s,1}^l|$ & -- & -- & -- \\  \hline
  $|\alpha_{s,2}^l|$ & $\sum\limits_{j=0,1,2}|\alpha_{s,i}^l|$ & -- & -- \\  \hline
  $|\alpha_{s,3}^l|$ & $\sum\limits_{j=0,1,3}|\alpha_{s,i}^l|$ & $\sum\limits_{j=0,2,3}|\alpha_{s,i}^l|$ & $\sum\limits_{j=1,2,3}|\alpha_{s,i}^l|$ \\  \hline
  $|\alpha_{s,4}^l|$ & $\sum\limits_{j=0,1,4}|\alpha_{s,i}^l|$ & $\cdots$ & $\cdots$ \\  \hline
  $\cdots$ & $\cdots$ & $\cdots$ & $\cdots$ \\  \hline
  $|\alpha_{s,7}^l|$ & $\cdots$ & $\cdots$ & $\cdots$ \\ \hline
\end{tabular}
}}
  \label{tab:SPC_partial_order}%
\end{table}%
As seen, any error patterns other than the those given in \emph{Proposition}~\ref{prop:SPC_node} will lead to more than 8 surviving paths with path metrics smaller than the 8-th path, which contradicts the assumption.
\end{IEEEproof}
\begin{remark}
According to Section~\ref{section:latency_analysis}, the latency (cycles) reduction from \cite{Dec:fast_SSCL} is $56 \rightarrow 15$ under $L=8$.
\end{remark}
\subsection{Error pattern identification via syndrome decoding}\label{section:syndrome_decoding}
Existing optimizations operate on special rates, e.g., R0/R1/SPC/Rep nodes. In this work, we suggest a parallelization method for arbitrary nodes with larger sizes (e.g., $B=8,16,\cdots$).

For general nodes, it is not easy to identify all possible error patterns as in R1/SPC nodes. However, it is possible to quickly narrow down to a subset of highly-likely error patterns for parallelized path extension. Syndrome decoding is particularly suitable here for two reasons, e.g., (i) blockwise syndrome calculation is simple and reuses the Kronecker product module, (ii) multiple error patterns (coset) can be pre-stored and retrieved in parallel. 
\subsubsection{General nodes}
\begin{figure*}
\centering
    \includegraphics[width= 0.8\textwidth]{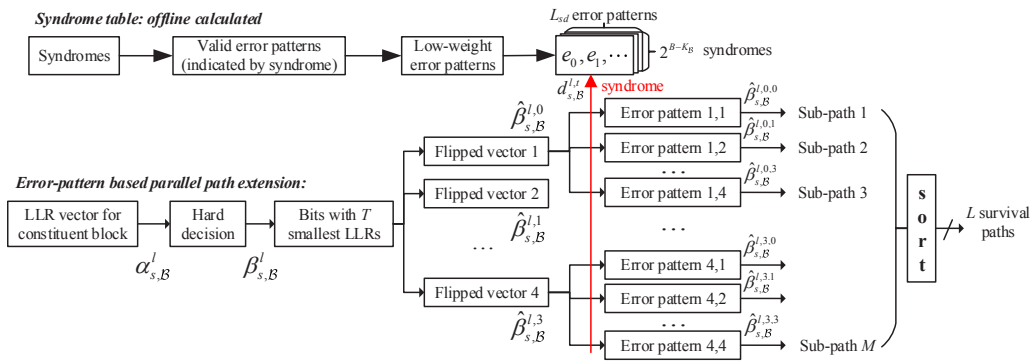}
    \caption{Error pattern identification via syndrome decoding.}
    \label{fig:syndrome_decoding}
\end{figure*}

As shown in Fig.~\ref{fig:syndrome_decoding}, we first obtain a set of input vectors via multi-bit hard decision and bit flipping. The flipped positions are chosen from the flipping set $\cal T$, i.e., the $T$ indices in ${\pmb \alpha}_{s,{\cal B}}$ with the smallest LLRs. Based on the hard estimation ${\pmb \beta}_{s,{\cal B}}^l$, we flip within $\cal T$ to generate $2^T$ input vectors, denoted by ${\pmb \beta}_{s,{\cal B}}^{l,t}$, and $t \in \{0\cdots 2^T-1\}$.

For example, if the $t$-th flipping pattern is ${\pmb e}_{i} \oplus {\pmb e}_{j} \oplus {\pmb e}_{k}$ (clearly $\{i,j,k\}\in {\cal T}$), then
\begin{equation}
{\pmb \beta}_{s,{\cal B}}^{l,t} = {\pmb \beta}_{s,{\cal B}}^l \oplus {\pmb e}_{i} \oplus {\pmb e}_{j} \oplus {\pmb e}_{k}.
\end{equation}

Given the flipping pattern, the syndrome-decoding-based parallel path extension is illustrated in Fig.~\ref{fig:syndrome_decoding}. The key steps, e.g., syndrome calculation and error pattern retrieval, are hardware-friendly binary operations and LUT.

Denote by $\textbf{G}_{\cal B} \triangleq \textbf{F}^{\otimes \log_2 B}$ the kernel of a general node and its frozen set $\cal F_B$, the parity-check matrix $\textbf{H}_{\cal B}$ is obtained by extracting the columns with indices in $\cal F_B$ from $\textbf{G}_{\cal B}$. Thus, the syndrome of vector ${\pmb \beta}_{s,{\cal B}}^{l,t}$ contains $B-K_{\cal B}$ bits and is calculated by
\begin{equation}
{\pmb d}_{s,{\cal B}}^{l,t} = \textbf{H}_{\cal B} \times {\pmb \beta}_{s,{\cal B}}^{l,t}.
\end{equation}

For each syndrome, its associated error patterns are computed offline \cite{book:ECC} and pre-stored by ascending weight order in LUT. Since a low-weight error pattern is more likely than a high-weight one, we only need to store a small number of lowest-weight patterns to reduce memory.

There are $2^{B-K_{\cal B}}$ different syndromes for a $(B,K_{\cal B})$ constituent code block, where $K_{\cal B}$ is the number of information bits within the block. As a result, the size of a syndrome table is $(2^{B-K_{\cal B}})\times L_{sd}$, where $L_{sd}$ is a constant number of error patterns pre-stored for each syndrome.

For example, the syndrome table for a general node with $B=8$, $K_{\cal B}=6$ and $L_{sd}=4$ has size $4 \times 4$ and is given in Table~\ref{tab:syndrome_table_B8_K6}.

\begin{table}
  \centering
  \caption{Syndrome table for $B=8, K_{\cal B}=6$}
{\footnotesize{
\begin{tabular}{|c|c|c|c|c|c|c|c|c|}
  \hline
  Syndrome & \multicolumn{4}{|c|}{Error Patterns (in Hex)} \\ \hline
  00 & 00 & 05 & 11 & 41 \\   \hline
  01 & 01 & 04 & 10 & 40 \\   \hline
  10 & 03 & 09 & 21 & 81 \\   \hline
  11 & 02 & 08 & 20 & 80 \\   \hline
\end{tabular}
}}
  \label{tab:syndrome_table_B8_K6}%
\end{table}%

The error patterns retrieved from LUT are used to simultaneously generate a set of candidate sub-paths, denoted by
\begin{equation}
\left\{\hat{\pmb \beta}_{s,{\cal B}}^{l,t,l_{sd}}\right\} = {\pmb \beta}_{s,{\cal B}}^{l,t} + \left\{error \ patterns \ \text{indexed by} \ {\pmb d}_{s,{\cal B}}^{l,t}\right\},
\end{equation}
where $t$ and $l_{sd}$ are the flipping pattern index and syndrome-wise error pattern index, respectively.

For each list path, we have $2^T\times L_{sd}$ extended sub-paths. The path metrics are updated according to \eqref{equ:multi_bit_PM_update} except that, the $T$ smallest LLRs are modified to a large value, i.e., $\alpha_{s,j}^{l,t} \rightarrow (-1)^{{\hat \beta}_{s,j}^{l,t}}\times\infty ,\forall j \in {\cal T}$, where ${\hat \beta}_{s,j}^{l,t}$ is the $j$-th hard bit after flipping. This procedure ensures at most one flip for each bit position and therefore no duplicate paths will survive, which is crucial to the overall performance.

Similar to R1/SPC nodes, the path extension and pruning is performed only one time for each block to keep $L$ surviving paths, i.e., ($L\rightarrow L\times2^T\times L_{sd}\rightarrow L$).

\begin{remark}\label{rmk:max_path_extension}
For small $K_{\cal B}$, an exhaustive-search-based path extension is more convenient since it generates $2^{K_{\cal B}}$ paths \cite{Dec:multi-bit}. For $K_{\cal B} > T+\log_2 {L_{sd}}$, it is more efficient to extend paths by the proposed flip-syndrome method. Therefore, we recommend to switch between exhaustive-search-based and syndrome-based path extension depending on the constituent code rate. As such, the maximum path extension is $\min\left(2^{K_{\cal B}},2^T\times L_{sd}\right)$.
\end{remark}
\begin{remark}
For a practical list size $L=8$, we can set $B=8, T\leq 2,L_{sd}\leq 4$ for 8-bit parallel decoding, or $B=16, T\leq 3,L_{sd}\leq 8$ for 16-bit parallel decoding to achieve a good tradeoff between complexity and latency, yet with negligible performance loss.
\end{remark}


\subsection{Latency analysis}\label{section:latency_analysis}
The minimum number of cycles is analyzed with the assumption that independent operations can be executed in parallel. In reality, the latency will be different depending on the number of processing elements available per implementation. However, the minimum cycle analysis represents the number of logical steps and provides a hardware-independent latency evaluation.

\begin{figure}
\centering
    \includegraphics[width= 0.5\textwidth]{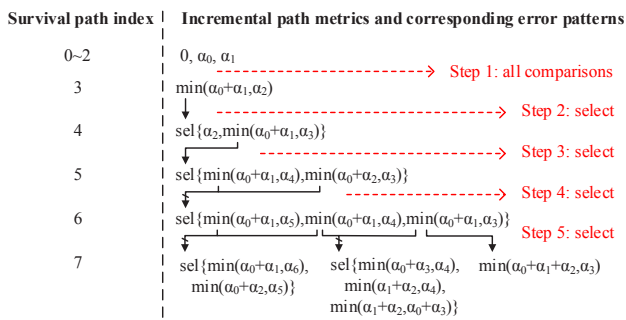}
    \caption{Minimum cycle analysis for a rate-1 node.}
    \label{fig:stage_analysis}
\end{figure}

For an R1 node, the 13 error patterns in \eqref{equ:rate_1_patterns} are retrieved from a pre-stored table, among which 8 are pre-selected according to path metric. The $13\rightarrow8$ path sorting and pruning logic is shown in Fig.~\ref{fig:stage_analysis}. For simplicity, $|\alpha_{s,t}^l|$ is abbreviated by $\alpha_t$. All relevant LLR pairs are compared in cycle 1. Among them, the first 3 pre-selected paths are ${\pmb \beta}_{s,{\cal B}}^l,{\pmb \beta}_{s,{\cal B}}^l\oplus {\pmb e}_{0}$ and ${\pmb \beta}_{s,{\cal B}}^l\oplus {\pmb e}_{1}$. The remaining paths are sequentially selected according to the comparison results and their preceding selection choices. Finally, the 8 candidate paths are pre-selected and sorted by ascending order. The process only requires 5 cycles.

Combining all sub-paths in an $L=8$ list decoder, there will be $8\times8=64$ paths for another round of pruning. Since the 8 sub-paths for each list are already ordered, the pruning requires an additional 9 cycles to identify the 8 survival paths \cite{Dec:bitonic}. The number of cycles are 14 and 15 for an R1 and SPC node, respectively.

For comparison, fast-SSCL \cite{Dec:fast_SSCL} requires 7 and 8 rounds of path extension and pruning for a Rate-1 and an SPC node, respectively. Each round takes a minimum of 7 cycles with bitonic sort \cite{Dec:bitonic}. Overall, a minimum of $7\times7=49$ and $7\times8=56$ cycles are required.

For general nodes, the proposed FSL decoder also has lower latency since more bits are processed in parallel. The overall latency is influenced by two factors (i) the number of leaf nodes in an SC decoding tree, (ii) the degree of parallelism within a leaf node.

For a rough estimation, the number of leaf nodes of a $N=1024,K=512$ Polar code is summarized in Table~\ref{tab:par_node_comparison}. The code is constructed by Polarization Weight (PW) \cite{Polar:PW}. For all schemes, the frozen bits before the first information bit are skipped. For R0/Rep/SPC/R1 nodes, the maximum length of a parallel processing block is $B_{\max}=32$. For general nodes, the parallel processing length is 8-bit or 16-bit, denoted by 8b and 16b FSL, respectively. As seen, 16b FSL only requires to visit a half of nodes to traverse the SC decoding tree.

\begin{table}
  \centering
  \caption{Number of leaf nodes in an SC decoding tree}
{\footnotesize{
\begin{tabular}{|c|c|c|c|c|c|c|c|}
  \hline
  Decoder & R0 & Rep & ML & Gen & SPC & R1 & Total\\ \hline
  4b ML \cite{Dec:multi-bit} & 30 & 0 & 154 & 0 & 0 & 0 & 184 \\ \hline
  F-SSCL \cite{Dec:fast_SSCL} & 21 & 23 & 0 & 0 & 23 & 24 & 91 \\ \hline
  8b FSL & 15 & 0 & 26 & 5 & 11 & 13 & 70 \\ \hline
  16b FSL & 7 & 0 & 14 & 11 & 5 & 9 & 46 \\ \hline
\end{tabular}
}}
  \label{tab:par_node_comparison}%
\end{table}%

To determine real latency, we synthesized the proposed decoders in TSMC 16nm CMOS with a frequency of 1GHz. The maximum supported code length is $N_{\max}=16384$, with LLRs and path metrics quantized to 6 bits. The number of processing elements is 128. The decoding latency of 4b multi-bit \cite{Dec:multi-bit}, Fast-SSCL \cite{Dec:fast_SSCL}, 8b FSL and 16b FSL decoders is measured at a code rate of $1/3$. For $N=1024$, the latency is 1258ns, 1079ns, 870ns and 697ns, respectively. For $N=4096$, the latency is 5134ns, 4239ns, 3640ns and 3003ns, respectively. The latency reduction from \cite{Dec:multi-bit,Dec:fast_SSCL} is $35\%\sim45\%$ and $29\%\sim42\%$, respectively. As seen, even compared with the most advanced SCL decoders \cite{Dec:multi-bit,Dec:fast_SSCL} in literature, the proposed 8b and 16b FSL decoders can further reduce latency. A detailed latency comparison is given in Table~\ref{tab:latency_comparison}.

\begin{table}
  \centering
  \caption{Comparison of decoding latency (ns)}
{\footnotesize{
\begin{tabular}{|c|c|c|c|c|c|}
  \hline
  \emph{N} & Rate & 4b ML \cite{Dec:multi-bit} & Fast-SSCL \cite{Dec:fast_SSCL} & 8b FSL & 16b FSL \\ \hline
  \multirow{2}{*}{1024} & $1/3$ & 1258 & 1079 & 870 & 697 \\ \cline{2-6}
  & $1/2$ & 1577 & 1307 & 1016 & 776 \\ \hline
  \multirow{2}{*}{4096} & $1/3$ & 5134 & 4239 & 3640 & 3003 \\ \cline{2-6}
  & $1/2$ & 6585 & 4984 & 4399 & 3501 \\ \hline
  \multirow{2}{*}{16384} & $1/3$ & 21717 & 17230 & 15879 & 13461 \\ \cline{2-6}
  & $1/2$ & 27357 & 19839 & 18763 & 15305 \\ \hline
\end{tabular}
}}
  \label{tab:latency_comparison}%
\end{table}%


\subsection{BLER performance}
\begin{figure}
\centering
    \includegraphics[width= 0.5\textwidth]{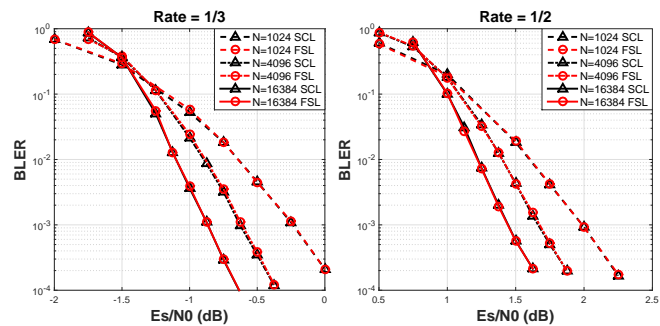}
    \caption{BLER comparison between SCL and FSL, both with $L=8$ and 16-bit CRC for final path selection.}
    \label{fig:BLER}
\end{figure}

The BLER performance of an FSL decoder is simulated and compared with its SCL decoder counterpart. For FSL, we adopt 16-bit parallel processing with $B=16, T\leq 3,L_{sd}\leq 8$. We simulated a wide range of code rates and lengths, and observe negligible performance loss. In the interest of space, only code rates $\{1/2,1/3\}$ and lengths $\{1024,4096,16384\}$ are plotted in Fig.~\ref{fig:BLER}. Throughout the paper, 16 CRC bits are appended to, but not included in, the $K$ payload bits. The code rate is calculated by $K/N$.

\section{Improved code construction}\label{section:construction}
Based on the proposed FSL decoder, we propose two code construction methods to further (i) reduce complexity and (ii) improve performance. The first method re-adjusts the information bit positions to avoid certain high-complexity constituent code blocks. The second one replaces outer constituent codes with optimized block codes to improve BLER performance.

\subsection{Adjusted Polar codes}
The complexity of an FSL decoder mainly arises from the size of syndrome tables. According to Section~\ref{section:syndrome_decoding}, the size of a syndrome table is $(2^{B-K_{\cal B}})\times L_{sd}$ for a constituent code block with $K_{\cal B}$ information bits and $L_{sd}$ error patterns per syndrome. According to Remark~\ref{rmk:max_path_extension}, a rate-dependent path extension is adopted, where the maximum path extension is $\min\left(2^{K_{\cal B}},2^T\times L_{sd}\right)$. In other words, high-complexity operations are incurred by medium-rate blocks, while high-rate or low-rate blocks can be processed with low complexity.

Thanks to the polarization effect, most blocks will diverge to high or low rates as code length increases, which is helpful. In the following, we show that, even for finite-length codes with insufficient polarization, we can deliberately eliminate some of the medium-rate blocks by re-adjusting their information bit positions.

\begin{figure}
\centering
    \includegraphics[width= 0.5\textwidth]{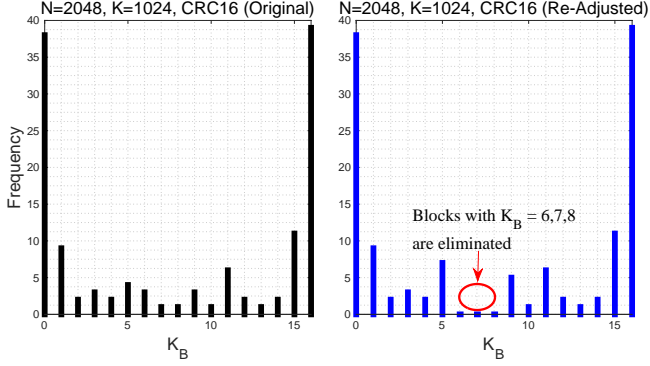}
    \caption{Block rate distribution before and after information bit re-adjustment.}
    \label{fig:hist_adj}
\end{figure}

For example, a 16-bit parallel FSL decoder with $B=16$, $T=3$ and $L_{sd}=8$ is used to decode a $N=2048, K=1024$, CRC16 Polar code. The original block rate distribution is shown on the left side of Fig.~\ref{fig:hist_adj}. As seen, many code blocks have already polarized to either high rate or low rate. Among the medium rate blocks, those with $K_{\cal B}=6$ are responsible for the majority decoding complexity (syndrome table size $1024$). However, there are only 3 such blocks. On the right side of Fig.~\ref{fig:hist_adj}, we eliminate blocks with $K_{\cal B}=6, 7$ and 8 by re-allocating their information bits to blocks with lower and higher rates. Although the adjusted Polar codes deviate from the actual polarization, which implies performance loss, they demand much lower decoding complexity. In particular, the largest syndrome table size reduces from $1024$ to $128$, with the information re-adjustment in Fig.~\ref{fig:hist_adj}.

Algorithm \ref{alg:rateAdjust} formalizes the above mentioned re-adjustment procedures. The input parameters are $K_{\cal B}^{low}$ and $K_{\cal B}^{high}$, which indicate that rates between $\frac {K_{\cal B}^{low}} B$ and $\frac {K_{\cal B}^{high}} B$ are to be eliminated. The algorithm first constructs an original Polar codes and determine the number of information bits $K_{\cal B}$ in each constituent code block. If a block has $K_{\cal B}^{low} < K_{\cal B} < K_{\cal B}^{high}$, the algorithm either adds or removes information bits within the block, until its rate $R_{\cal B}$ satisfies $R_{\cal B}\geq\frac {K_{\cal B}^{high}} B$ or $R_{\cal B}\leq\frac {K_{\cal B}^{low}} B$. Once the rate of a block is adjusted, another block has to change its rate accordingly to ensure that overall code rate remains unchanged.


\begin{algorithm}
\begin{algorithmic}
\STATE Input: $N,K,B,{\cal I},K_{\cal B}^{low}, K_{\cal B}^{high}$; Output: ${\cal I}_{adj}$
\STATE 1) Re-adjust to eliminate medium-rate block.
\FOR {each block with $K_{\cal B}^{low} < K_{\cal B} < K_{\cal B}^{high}$}
\IF {$K_{\cal B} - K_{\cal B}^{low} < K_{\cal B}^{high} - K_{\cal B}$}
\STATE Reduce $K_{\cal B}$ to $K'_{\cal B} = K_{\cal B}^{low}$
\ELSE
\STATE Increase $K_{\cal B}$ to $K'_{\cal B} = K_{\cal B}^{high}$
\ENDIF
\ENDFOR
\STATE 2) Balance overall rate when necessary.
\FOR {each block with $K'_{\cal B} = K_{\cal B}^{low}$ (or $K'_{\cal B} = K_{\cal B}^{high}$)}
\WHILE {Total $\#$ info. bits $\sum {K'_{\cal B}} > K$ (or $< K$)}
\STATE Reduce $K'_{\cal B}$ to $K'_{\cal B} = K_{\cal B}^{low}-1$
\STATE (or Increase $K'_{\cal B}$ to $K'_{\cal B} = K_{\cal B}^{high}+1$)
\ENDWHILE
\ENDFOR
\STATE 3) Select information bits.
\FOR {each constituent code block}
\STATE Select $K'_{\cal B}$ most reliable bit positions to ${\cal I}_{adj}$
\ENDFOR
\end{algorithmic}
\caption{An information bit re-adjustment algorithm}
\label{alg:rateAdjust}
\end{algorithm}

\begin{figure}
\centering
    \includegraphics[width=0.5\textwidth]{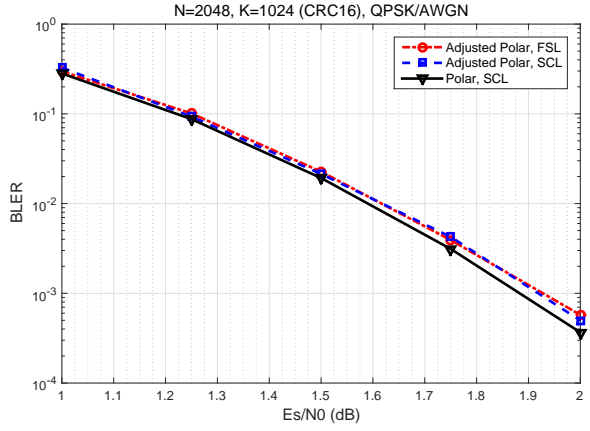}
    \caption{Comparison between Polar codes under SCL decoder ($L=8$) and Adjusted-Polar codes under both SCL and FSL decoders ($L=8, B=16$) with code length $N=2048$.}
    \label{fig:N2048_adjust}
\end{figure}

\begin{figure}
\centering
    \includegraphics[width=0.5\textwidth]{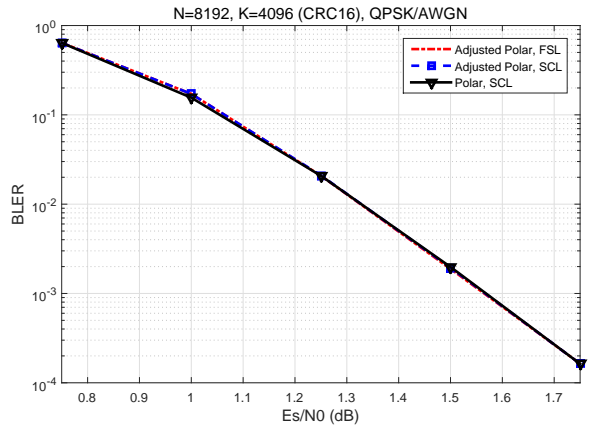}
    \caption{Comparison between Polar codes under SCL decoder ($L=8$) and Adjusted-Polar codes under both SCL and FSL decoders ($L=8, B=16$) with code length $N=8192$.}
    \label{fig:N8192_adjust}
\end{figure}

Fig.~\ref{fig:N2048_adjust} shows the performance of $N=2048, K=1024$ Polar codes and Adjusted Polar codes under both SCL and FSL decoders with $L=8$. For Adjusted Polar codes, a 16b FSL decoder ($B=16$) is implemented, and blocks with $K_{\cal B}=6, 7$ and 8 are eliminated. The syndrome table size thus reduces from $1024$ to $128$. The BLER loss due to information bit re-adjustment is only 0.02dB at BLER $1\%$. The same experiment is conducted for $N=8192, K=4096$, whereas the performance loss becomes negligible as shown in Fig.~\ref{fig:N8192_adjust}. This can be well explained: medium-rate blocks reduce as polarization increases with code length, thus requiring less re-adjustment and incurring less performance loss.

The proposed construction allows us to trade some performance for significant complexity reduction, thus bears practical importance.

\subsection{Optimized outer codes}
Observe that the proposed hard-input decoder for outer block codes is no longer an SC decoder, but similar to an ML decoder. However, the default Polar outer codes have poor minimum distance and may not be suitable for the proposed decoder. To obtain a better performance, a straightforward idea is to adopt outer codes with optimized distance spectrum.

Note that the error-pattern-based decoders do not need to change at all. As long as the generator/parity-check matrices are defined, the outer decoders only need to update the error patterns according to that specific code. That means any linear block codes fit well into the FSL decoding framework, offering full freedom to optimize the outer codes.

For $B=16$, we present a specific outer code design for each $K_{\cal B}$. For example, $K=2$ simplex codes repeated to length-16 have a minimum distance 10, which is larger than 8 of $(16,2)$ Polar codes. Following this idea, we individually optimize each $(B,K_{\cal B})$ outer codes with respect to code distance.

For $K_{\cal B}=2,3,4$, repetition over simplex codes always yields higher code distance than the corresponding Polar codes. Their respective generator matrices $G_{K_{\cal B}}$ are
\begin{equation*}
G_2 = \left[S_2 \ S_2 \ S_2 \ S_2 \ S_2 \ \begin{matrix}
1\\
1
\end{matrix} \right], \ S_2= \left[\begin{matrix}
1 \ 1 \ 0\\
1 \ 0 \ 1
\end{matrix} \right];
\end{equation*}
\begin{equation*}
G_3 = \left[S_3 \ S_3 \ \begin{matrix}
1 \ 1\\
1 \ 1\\
1 \ 0
\end{matrix} \right], \ S_3= \left[\begin{matrix}
1 \ 1 \ 1 \ 1 \ 0 \ 0 \ 0\\
1 \ 1 \ 0 \ 0 \ 1 \ 1 \ 0\\
1 \ 0 \ 1 \ 0 \ 1 \ 0 \ 1
\end{matrix} \right];
\end{equation*}
\begin{equation*}
G_4 = \left[S_4 \ \begin{matrix}
1\\
1\\
1\\
1
\end{matrix} \right], \ S_4= \left[\begin{matrix}
1 \ 1 \ 1 \ 1 \ 1 \ 1 \ 1 \ 1 \ 0 \ 0 \ 0 \ 0 \ 0 \ 0 \ 0 \\
1 \ 1 \ 1 \ 1 \ 0 \ 0 \ 0 \ 0 \ 1 \ 1 \ 1 \ 1 \ 0 \ 0 \ 0 \\
1 \ 1 \ 0 \ 0 \ 1 \ 1 \ 0 \ 0 \ 1 \ 1 \ 0 \ 0 \ 1 \ 1 \ 0 \\
1 \ 0 \ 1 \ 0 \ 1 \ 0 \ 1 \ 0 \ 1 \ 0 \ 1 \ 0 \ 1 \ 0 \ 1
\end{matrix} \right].
\end{equation*}

For $K_{\cal B}=6,7$, extended BCH (eBCH) codes also yields better distance spectrum than the corresponding Polar codes. Their respective generator matrices $G_{K_{\cal B}}$ are
\begin{equation*}
G_6 = \left[\begin{matrix}
0 \ 0 \ 1 \ 0 \ 0 \ 1 \ 0 \ 0 \ 1 \ 1 \ 1 \ 0 \ 1 \ 0 \ 0 \ 0 \\
1 \ 1 \ 1 \ 1 \ 1 \ 1 \ 1 \ 1 \ 0 \ 0 \ 0 \ 0 \ 0 \ 0 \ 0 \ 0 \\
1 \ 1 \ 1 \ 1 \ 0 \ 0 \ 0 \ 0 \ 1 \ 1 \ 1 \ 1 \ 0 \ 0 \ 0 \ 0 \\
1 \ 1 \ 0 \ 0 \ 1 \ 1 \ 0 \ 0 \ 1 \ 1 \ 0 \ 0 \ 1 \ 1 \ 0 \ 0 \\
1 \ 0 \ 1 \ 0 \ 1 \ 0 \ 1 \ 0 \ 1 \ 0 \ 1 \ 0 \ 1 \ 0 \ 1 \ 0 \\
1 \ 1 \ 1 \ 1 \ 1 \ 1 \ 1 \ 1 \ 1 \ 1 \ 1 \ 1 \ 1 \ 1 \ 1 \ 1
\end{matrix} \right];
\end{equation*}
\begin{equation*}
G_7 = \left[\begin{matrix}
0 \ 1 \ 1 \ 1 \ 0 \ 0 \ 1 \ 0 \ 0 \ 0 \ 1 \ 0 \ 1 \ 0 \ 0 \ 0 \\
0 \ 0 \ 1 \ 0 \ 0 \ 1 \ 0 \ 0 \ 1 \ 1 \ 1 \ 0 \ 1 \ 0 \ 0 \ 0 \\
1 \ 1 \ 1 \ 1 \ 1 \ 1 \ 1 \ 1 \ 0 \ 0 \ 0 \ 0 \ 0 \ 0 \ 0 \ 0 \\
1 \ 1 \ 1 \ 1 \ 0 \ 0 \ 0 \ 0 \ 1 \ 1 \ 1 \ 1 \ 0 \ 0 \ 0 \ 0 \\
1 \ 1 \ 0 \ 0 \ 1 \ 1 \ 0 \ 0 \ 1 \ 1 \ 0 \ 0 \ 1 \ 1 \ 0 \ 0 \\
1 \ 0 \ 1 \ 0 \ 1 \ 0 \ 1 \ 0 \ 1 \ 0 \ 1 \ 0 \ 1 \ 0 \ 1 \ 0 \\
1 \ 1 \ 1 \ 1 \ 1 \ 1 \ 1 \ 1 \ 1 \ 1 \ 1 \ 1 \ 1 \ 1 \ 1 \ 1
\end{matrix} \right].
\end{equation*}

For $K_{\cal B}=8,9$, the dual of eBCH codes are adopted; for $K_{\cal B}=12,13,14$, the dual of simplex codes are adopted. For the remaining rates, the original Polar codes are adopted.

Depending on $K_{\cal B}$, the outer codes are combination of different codes, or hybrid outer codes. The resulting concatenated codes are thus called \emph{hybrid-Polar codes}. Note that the lengths of the outer codes are not necessarily power of 2, making the concatenated codes length compatible.

\begin{figure}
\centering
    \includegraphics[width= 0.45\textwidth]{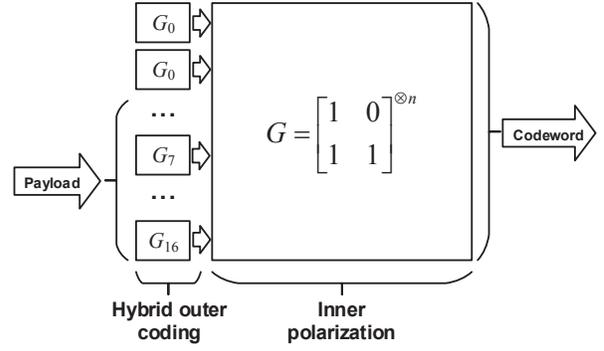}
    \caption{A Hybrid-Polar encoding flow.}
    \label{fig:Hybrid-Polar-encoding}
\end{figure}

The encoding steps are shown in Fig.~\ref{fig:Hybrid-Polar-encoding}, and explained as follows:
\begin{enumerate}
  \item First, an original $(N,K)$ Polar code is constructed, in order to determine the rate of each $(B,K_{\cal B})$ outer code.
  \item Second, each block is individually encoded, i.e., multiplying a length-$K_{\cal B}$ information vector by the corresponding generator matrix.
  \item Thirdly, the outer code words are concatenated into a long intermediate vector, upon which inner polarization is performed to obtain a single code word.
\end{enumerate}

The proposed outer codes have better distance spectrum than Polar codes. The code weights $\{w\}$ are enumerated in Table~\ref{tab:distSpectrum}. The numbers of code words having a specific weight are displayed, and those of minimum weight are highlighted in boldface. As seen, the distance spectrum of the hybrid codes improves upon Polar codes with the same $K_{\cal B}$ in two ways:
\begin{itemize}
  \item The minimum distance increases, e.g., $K_{\cal B}=2,6,7$;
  \item The minimum distance remains the same, but the number of minimum-weight codewords reduces, e.g., $K_{\cal B}=3,4,9,10,12,13,14$.
\end{itemize}

\begin{table*}[t]
  \centering
  \caption{Comparison of distance spectrum between outer codes: Polar vs Hybrid}
    \begin{tabular}{|c|c|c|c|c|c|c|c|c|c|c|c|c|c|c|c|c|c|}
    \hline
    Code & $K_{\cal B}, \{w\}$ & 1 & 2 & 3 & 4 & 5 & 6 & 7 & 8 & 9 & 10 & 11 & 12 & 13 & 14 & 15 & 16 \\
    \hline
    Polar & 2 & 0 & 0 & 0 & 0 & 0 & 0 & 0 & \textbf{2} & 0 & 0 & 0 & 0 & 0 & 0 & 0 & 1 \\
    \hline
    Simplex & 2 & 0 & 0 & 0 & 0 & 0 & 0 & 0 & 0 & 0 & \textbf{1} & 2 & 0 & 0 & 0 & 0 & 0 \\
    \hline
    Polar & 3 & 0 & 0 & 0 & 0 & 0 & 0 & 0 & \textbf{6} & 0 & 0 & 0 & 0 & 0 & 0 & 0 & 1 \\
    \hline
    Simplex & 3 & 0 & 0 & 0 & 0 & 0 & 0 & 0 & \textbf{1} & 4 & 2 & 0 & 0 & 0 & 0 & 0 & 0 \\
    \hline
    Polar & 4 & 0 & 0 & 0 & 0 & 0 & 0 & 0 & \textbf{14} & 0 & 0 & 0 & 0 & 0 & 0 & 0 & 1 \\
    \hline
    Simplex & 4 & 0 & 0 & 0 & 0 & 0 & 0 & 0 & \textbf{7} & 8 & 0 & 0 & 0 & 0 & 0 & 0 & 0 \\
    \hline
    Polar & 5 & 0 & 0 & 0 & 0 & 0 & 0 & 0 & 30 & 0 & 0 & 0 & 0 & 0 & 0 & 0 & 1 \\
    \hline
    Polar & 6 & 0 & 0 & 0 & \textbf{4} & 0 & 0 & 0 & 54 & 0 & 0 & 0 & 4 & 0 & 0 & 0 & 1 \\
    \hline
    eBCH & 6 & 0 & 0 & 0 & 0 & 0 & \textbf{16} & 0 & 30 & 0 & 16 & 0 & 0 & 0 & 0 & 0 & 1 \\
    \hline
    Polar & 7 & 0 & 0 & 0 & \textbf{12} & 0 & 0 & 0 & 102 & 0 & 0 & 0 & 12 & 0 & 0 & 0 & 1 \\
    \hline
    eBCH & 7 & 0 & 0 & 0 & 0 & 0 & \textbf{48} & 0 & 30 & 0 & 48 & 0 & 0 & 0 & 0 & 0 & 1 \\
    \hline
    Polar & 8 & 0 & 0 & 0 & 28 & 0 & 0 & 0 & 198 & 0 & 0 & 0 & 28 & 0 & 0 & 0 & 1 \\
    \hline
    Polar & 9 & 0 & 0 & 0 & \textbf{44} & 0 & 64 & 0 & 294 & 0 & 64 & 0 & 44 & 0 & 0 & 0 & 1 \\
    \hline
    Dual of eBCH & 9 & 0 & 0 & 0 & \textbf{20} & 0 & 160 & 0 & 150 & 0 & 160 & 0 & 20 & 0 & 0 & 0 & 1 \\
    \hline
    Polar & 10 & 0 & 0 & 0 & \textbf{76} & 0 & 192 & 0 & 486 & 0 & 192 & 0 & 76 & 0 & 0 & 0 & 1 \\
    \hline
    Dual of eBCH & 10 & 0 & 0 & 0 & \textbf{60} & 0 & 256 & 0 & 390 & 0 & 256 & 0 & 60 & 0 & 0 & 0 & 1 \\
    \hline
    Polar & 11 & 0 & 0 & 0 & 140 & 0 & 448 & 0 & 870 & 0 & 448 & 0 & 140 & 0 & 0 & 0 & 1 \\
    \hline
    Polar & 12 & 0 & \textbf{8} & 0 & 252 & 0 & 952 & 0 & 1670 & 0 & 952 & 0 & 252 & 0 & 8 & 0 & 1 \\
    \hline
    Dual of Simplex & 12 & 0 & \textbf{1} & 42 & 133 & 252 & 469 & 750 & 835 & 680 & 483 & 294 & 119 & 28 & 7 & 2 & 0 \\
    \hline
    Polar & 13 & 0 & \textbf{24} & 0 & 476 & 0 & 1960 & 0 & 3270 & 0 & 1960 & 0 & 476 & 0 & 24 & 0 & 1 \\
    \hline
    Dual of Simplex & 13 & 0 & \textbf{11} & 82 & 233 & 516 & 1003 & 1470 & 1595 & 1400 & 1017 & 558 & 219 & 68 & 17 & 2 & 0 \\
    \hline
    Polar & 14 & 0 & \textbf{56} & 0 & 924 & 0 & 3976 & 0 & 6470 & 0 & 3976 & 0 & 924 & 0 & 56 & 0 & 1 \\
    \hline
    Dual of Simplex & 14 & 0 & \textbf{35} & 150 & 425 & 1100 & 2051 & 2810 & 3195 & 2920 & 1985 & 1066 & 475 & 140 & 25 & 6 & 0 \\
    \hline
    Polar & 15 & 0 & 120 & 0 & 1820 & 0 & 8008 & 0 & 12870 & 0 & 8008 & 0 & 1820 & 0 & 120 & 0 & 1 \\
    \hline
    \end{tabular}%
  \label{tab:distSpectrum}%
\end{table*}%

\begin{figure}
\centering
    \includegraphics[width= 0.5\textwidth]{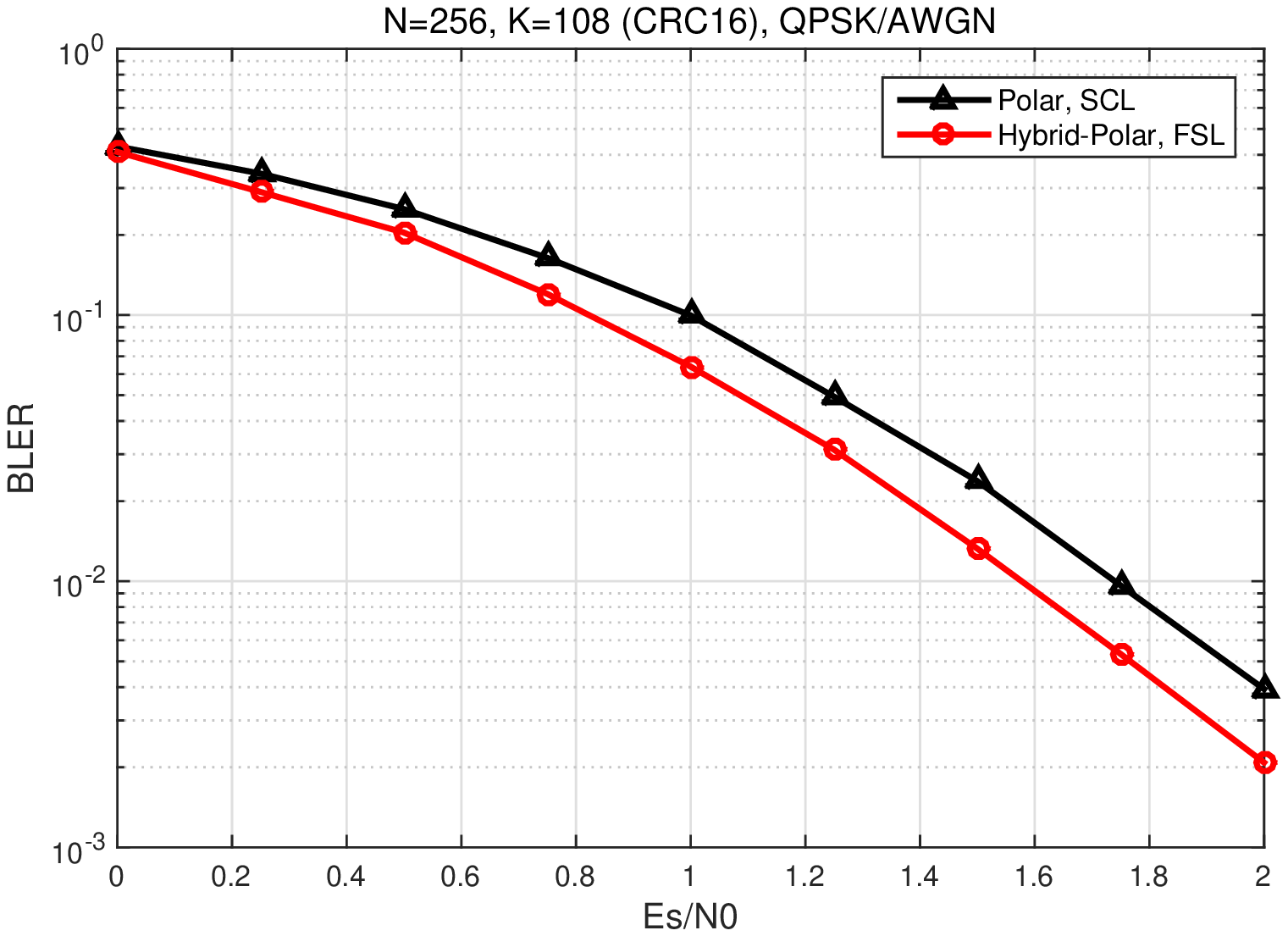}
    \caption{Comparison between Polar codes under SCL decoder ($L=8$) and Hybrid-Polar codes under FSL decoder ($L=8, B=16$) with code length $N=256$.}
    \label{fig:Hybrid256}
\end{figure}

\begin{figure}
\centering
    \includegraphics[width= 0.5\textwidth]{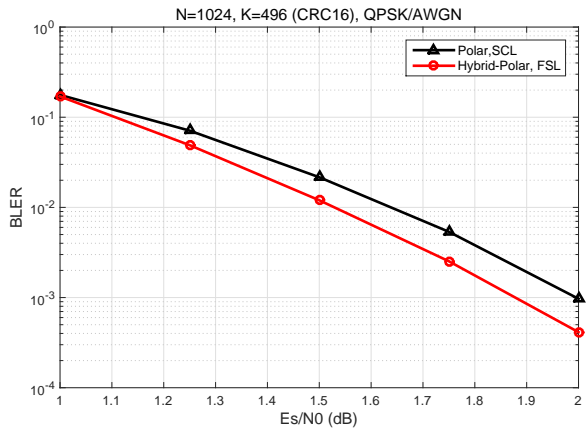}
    \caption{Comparison between Polar codes under SCL decoder ($L=8$) and Hybrid-Polar codes under FSL decoder ($L=8, B=16$) with code length $N=1024$.}
    \label{fig:Hybrid1024}
\end{figure}

Fig.~\ref{fig:Hybrid256} and Fig.~\ref{fig:Hybrid1024} show the performance of $N=256$ and $N=1024$ Polar codes, respectively, along with hybrid-Polar codes of the same length and rate. As seen, a performance gain between $0.1 \sim 0.2$ dB is demonstrated.

Since such BLER improvement comes with no additional cost within the FSL decoder architecture, the Hybrid-Polar codes is considered worthwhile in practical implementations.

\section{Conclusions}\label{section:conclusion}
In this work, we propose the hardware architecture of a flip-syndrome-list decoder to reduce decoding latency with improved parallelism. A limited number of error patterns are pre-stored, and simultaneously retrieved for bit-flipping-based path extension. For R1 and SPC nodes, only 13 error patterns are pre-stored with no performance loss under list size $L=8$; for general nodes, we may further reduce latency with a syndrome table to quickly identify a set of highly likely error patterns. Based on the decoder, two code construction optimizations are proposed to either further reduce complexity or improve performance. The proposed decoder architecture and code construction are designed particularly for applications with low-latency requirements.

%

\ifCLASSOPTIONcaptionsoff
  \newpage
\fi

\end{document}